**Long Paper**

# Level of Awareness of PSU – Bayambang Campus Students towards E–learning Technologies

Matthew John F. Sino Cruz
Information Technology Department, Pangasinan State University - Bayambang Campus, Philippines
zengmat@gmail.com (corresponding author)

Kim Eric B. Nanlabi
Information Technology Department, Pangasinan State University - Bayambang Campus, Philippines
kimericbnanlabi@psu.edu.ph

Michael Ryan C. Peoro
Information Technology Department, Pangasinan State University - Bayambang Campus, Philippines
mike.peoro@yahoo.com

## Abstract

*Purpose* – The study determines and measures the level of awareness of the PSU – Bayambang Campus students towards different e – learning technologies.

*Method* – The researchers employed Quantitative Research Approach. The study determined the profile of the respondents through a demographic questionnaires and the current status of the campus in terms of the ICT Resources and Network Infrastructure.

*Result* – The survey was carried out in order to measure the level of awareness towards different e – learning technologies among students. It was measured in terms of their familiarity to the existing e – learning technologies as well as the known features of these technologies. Although 52.50% of the respondents seem to be familiar to the concepts of e – learning, it is important to take into consideration the extent of their exposure to these technologies and its level of utilization in order to support the learning process.

Conclusion – Technology, Support and Users were considered to be important factors that affect the awareness of the students. These technologies can be used to improve the existing learning process if there is enough support for its implementation through policies and provision of the needed resources.

Recommendation – In order to improve the awareness of the stakeholders, the researchers recommended that policies on the integration of e – learning technologies must be designed and integrated to the existing learning process. The administration should also provide the needed ICT Resources and Infrastructure in order to support the use of these technologies. They should also provide training for the students and teachers in order to improve their awareness and enable them to use and maximize the benefits of these technologies.

Research Implication – The research can serve as a guide for the design of the policies since it provide an overview of the current status of the university in terms of awareness of the students towards these technologies. These technologies can be used to improve the existing learning process of the University and provide a wider avenue for learning and interaction.

---

## INTRODUCTION

Higher education often talks about the expectations of today's generation in using technology in a typical learning environment (Roberts, 2018). Through the application of Information Technology, today's learning environment or spaces have the potential to serve the new learning environment (Brown, 2005). In view with this, learning as well as the space where it takes place is considered with utmost importance.

The resources used in higher education are increasingly electronic and delivered over the network. This approach has realized the goal of achieving fully ubiquitous access to the resources available on the internet. The digitization of these resources paved the way to the integration of the learning space with technology (Brown, 2005). It was observed that today's generation desire for the use of technology to support learning and expect that professionals will use it as a medium to better communicate knowledge (Roberts, 2018). This paradigm shift has expanded and evolved the nature of classroom and brought about an interest in new pedagogical approaches in learning process. The phenomenon has given a way to a space / environment wherein learning can take place regardless of the time and location. This environment should accommodate the use of as many kinds of materials as possible and enable the display of and access to these materials by all participants (Brown, 2005). One of the most popular terms used to describe educational process over virtual space is e – learning (Simpsons, 2005).

The successful implementation of e-learning depends on different factors that will affect the delivery of knowledge towards the learners. As noted by Conole (2010), these factors include socio – cultural factor, methodology and supportive technologies. This was supported by Jefferies and McRobb (2007) and Stahl (2004) who stated that e – learning is composed of three major domains namely pedagogy, ethics and technology. E – Learning is supported by the different forms of technology which bridge the distance between the learner and the teacher. It depends on the hardware, software, communication media, applications, course management, computer – mediated communication and virtual environment for its implementation (Haythornthwaite et al., 2007).

Pangasinan State University is one of the State Universities in the Philippines which gives access to the poor but deserving students with quality education. In doing so, the current administration has redesigned its Institutional Outcome and concentrated in providing excellent academic service to its clientele, who are the students. Guided by this principle, the researcher has seen the opportunity to integrate the use of modern technology in improving the services provided to the students. However, certain factors should be considered before totally integrating technology with the existing learning process in the campus and determine whether or not it will be fully materialized in the long run. Moreover, this paper tries to determine the level of awareness of the existing learning management systems available in the internet among the students and faculty members of the campus. It will be done by considering the factors identified in the Critical Success Factors Model. The model will guide the researchers to determine what are the issues in terms of usability are in the existing learning management systems available over the internet and provide recommendations on how to incorporate these technologies in the learning process in the campus.

## Objectives

The study aims to measure the level of awareness of the students and faculty members of PSU – Bayambang Campus towards e – learning technologies (e.g. Massively Open Online Courses (MOOCs), Learning Management System (LMS), Blended Learning,

Open University System, Multimedia Courseware and Webcast Technologies)and design a framework for integrating LMS to the current learning practices in the Campus.

Hence, it specifically seeks to:

1. describe the profile of the students in terms of:
   a. Age
   b. Gender
   c. Program / Degree
   d. Year Level
   e. Academic Status
2. describe the profile of the campus in terms of:
   a. No. of serviceable computer units intended for internet usage
   b. Network infrastructure
3. determine the level of awareness of the students towards different e – learning technologies

## LITERATURE REVIEW

### *Gamification in the Context of Learning Process*

Learning process is indeed a two – way recursive process which involves teaching and learning on both the learner and the teacher. Technology has influenced the rapid paradigm shift by incorporating several tools that can facilitate and support these processes. Several studies were conducted that shows the use of different technological interventions in improving the learning process in different context.

One area in which technology is seen as a means for improving the learning process is Gamification. Gamification is the application of game components in supporting the different elements of the learning process (Kulpa, 2017). Some of the popular game components that can be used include animation, rewards, challenges and competitions, avatars and progress (Kostromina & Gnedykh, 2016; Domínguez et al., 2013; Kulpa, 2017; Baxter, Holderness & Wood, 2016). Studies have shown that these components have great effects on student's motivations towards the usage of the e – learning environment (Domínguez et al., 2013). The information presented visually using these components is shown to be influential for the learning process of the students (Kostromina & Gnedykh, 2016; Loughrey & O'Broin, 2017). However, it is also pointed out that aside from using the known components of the game in the learning process, the efficient usage of e – learning environment also depends on the self – discipline of the learners (Gorbunovs, Kapenieks & Cakula, 2016). The effect of gamification towards student's performance relies on the extent to which the learners engage in the different activity in the learning environment and how they use the instructional materials available in the medium (Hans, 2015; Giannetto, Chao & Fontana, 2013).

The components identified are then implemented using different activities and features of the gamified learning environment. For content – related activity, the environment may incorporate multimedia presentation and overall class activities such as accessing instructional materials and resources (Kostromina & Gnedykh, 2016). On the other hand, task – related activity includes Task Completion, Leaderboards and Progress Reports (Domínguez et al., 2013; Kulpa, 2017; McGuire et al., 2017; Paisley, 2013; Furdu, Tomozei & Kose, 2017; Olsson, Mozelius & Collin, 2015). Most of the games are self–paced. It means that the learners are in control of their own learning using the medium and they are the one who decide when they are going to access the materials and accomplish the task (Baxter et al., 2016; Loughrey & O'Broin, 2017; Lambert, 2017; Pais, Nogues & Munoz, 2017; Olsson et al., 2015).

Even though the mode of learning is remote and distant between each learners and teachers, gamified learning environment still foster the essence of team work and competition among the learners. Just like in a typical game, competition arises whenever learner tries to compete with other learners in accomplishing tasks using the medium (Giannetto et al., 2013). On the other hand, socialization is likewise fostered in these learning environments. Collaborative Learning is implemented by providing the ability to communicate with other learners as well as with the teacher. It may be in the form of forum, feedback mechanisms or private messaging components which are very helpful to establish an active interaction between these actors.

## Application of Multimedia – Based Learning Systems

Many literature and publications were published which tackles the effective use of multimedia in providing quality education to the students. According to Chen, Wang and Wu (2009), multimedia courseware is the effective use of the different forms of multimedia elements in presenting or displaying the information to the users. Thus, the courseware provides a more interactive way on delivering instructions to the students.

Likewise, Computer Based Training was used by the military offices in US to conduct military training. Thus, it was proven that the Computer Based Training was a lot more effective on delivering instruction than the standard classroom – lecture session. The development of the computer based instruction continuous and escalates with the utilization of the internet. The scenario led to the development of the so – called online learning (Childress, 2017).

With the development and popularity of modern computers and internet, Web – based training allows users to take the course with nominal cost yet, enjoying the advantages provided by the training. The article also enumerated the advantages of engaging to a web – based training over the traditional one. The main advantage of web – based training is that, the materials are easier and quicker to update for as long as you have an internet connection and interaction with other online learners.

Other author such as Abu Bakar, Ismail and Zainal Abidin (2015) cited the benefits of using Multimedia Courseware in delivering education. According to the author, interactive multimedia takes less time to implement, but is enjoyed more and increases learning. He also cited that learning was higher when information was presented via computer-based multimedia systems than traditional classroom lectures. This was proven since most students were engaged to continue using the courseware, for it provides a different approach on delivering the lessons to its intended users.

In the like manner, an article written by an unknown writer cited the difference between video lecturing and multimedia form of learning. The author stated that video is linear by nature which means that the user needs to follow series of sequence before they can proceed to the topic they are interested. However, in multimedia courseware, the users were given enough navigational elements which enable them to proceed to the topic they want even without having the need to following a specific sequence of topics. Thus, with this capability, the multimedia courseware falls under nonlinear category which means that the users can take full control over the flow of the courseware (Abu Bakar et al., 2015).

Aside from the literature that were published which tackle more on the evolution, uses and benefits of the multimedia courseware, there are also studies which discussed the effectiveness of the courseware on the learning process of the students. One of the authors who conducted a study similar to the current study was Ibrahim, Krishnasamy and Abdullah (1996). The study focused more on the factors concerning the development and usage of multimedia courseware in the academe. He noted on his study that before implementing multimedia courseware, the instructors should have formulated a well – thought approached in order to incorporate it properly to regular class discussions. He also concluded that the graphics layout, audio, text orientation and design should be well furnished for they become an essential factor on the effectiveness of the multimedia courseware.

Zaman et al. (2000) also conducted a study on the effectiveness of the multimedia courseware on the learning process of the preschool students. The result of her study shows that the students were motivated to learn more about the topic since the courseware offers intensive interactivity and multimedia elements in presenting the topic. Her study also shows that the learning capacity of the students is much faster compare to the traditional classroom – based discussion.

Liu (2010) conducted a study entitled "An Experimental Study on the Effectiveness of Multimedia in College English Teaching" to explore the effectiveness of multimedia assisted methods in College English Teaching. After conducting the study, she concluded that the current multi-media assisted teaching method does not facilitate a two-way communication atmosphere, student-oriented classroom, or cultivating students' independent learning ability. She pointed out that the most efficient way to which English teachers could utilize the use of courseware in teaching the subject is with the aid of an audio – visual input on presenting and organizing the materials.

## Factors Affecting the Awareness and Acceptance of E – Learning Technologies

Studies about the level of awareness among stakeholders have been conducted in different areas. A study conducted by Edumadze et al. (2014) reveals that many lecturers fail to use e – learning tools because they are not proficient enough in using them. They seem to be aware of e – learning tools, such as LMS, since they have heard the term but do not have any experience in using the technology. The study shows that the primary factors for the successful implementation of e – learning technology include skills, access to the internet and an environment to support the technology. Having the facility to support the technology may lead to the willingness of the stakeholders to adopt the technology. The findings was supported by the study of Agboola (2006) who stated that basic computer skills and the ability to access the internet and the services on it are some of the primary factors for the successful implementation of e – learning. He also added that e – learning training is the best predictor for e – learning adoption and readiness. It is necessary to educate the stakeholders about what constitutes e – learning since it will determine the extent to which they can use the technology and gain experience (Ngampornchai & Adams, 2016; Almohod & Shafi, 2013). This is due to the fact that there are still unaware and misconceptions of what constitutes e–learning. The study of Edumadze et al. (2014) shows that stakeholders often have a very limited understanding about e-learning concept. They usually thought of it as using PowerPoint Presentation, Chat–based discussion and involvement in a discussion forum. Hence, users need to be well educated and oriented in order for them to have a better grasp about the e–learning (Agboola, 2006).

The acceptance of the technology also relies on the attitude of the stakeholders of the institution. A positive attitude leads to the greater possibility of using and maximizing the e – learning tools (Kar, Saha & Mondal, 2014; Fitzpatrick, 2012; Almohod & Shafi, 2013). The willingness to adopt and implement e – learning technology is greatly affected by acquiring skills required through professional training and other technical support (Agboola, 2006). Gaining knowledge and skills about these technologies may likely affect their engagement on the different tools that e–learning technologies provide (Olibie, Ezoem & Ekene, 2014).

Other studies suggest that the availability of ICT resources also affects the implementation and acceptance of e – learning technology (Ngampornchai & Adams, 2016; Fitzpatrick, 2012). Even though the stakeholders have the skills in using computers and utilizing internet resources, it is also important to consider the availability of the resources that needs to support the implementation of the e – learning technology (Edumadze et al., 2014; Olibie et al., 2014). Having the needed resources will enable the teacher and the learner utilize the technology and enjoy its benefits. However, some still argue that even though the users are familiar with the technology or have the necessary resources, they may or may not have a positive perception toward the technology (Ngampornchai & Adams, 2016).

It was also suggested that the organization's commitment and management support to implement e – learning technology are also primary factors for its acceptance among its users. If the management and organization do not show its support for the initiative, then the users may not accept the technology and the initiative will not be successful (Almohod & Shafi, 2013; Fitzpatrick, 2012).

Fitzpatrick (2012) also added that Marketing, Training Design, Support and Structure are also essential factors for the successful implementation and adoption of e – learning technologies. The quality of course content, transparency in assessment and evaluation, collaboration and industry acceptance are also important components that leads to a design of an effective e – learning technology (Agariya & Singh, 2012). With this factors observed, e–learning is seen as a powerful way of enhancing and learning. Moreover, it is still important to note that the development of e–learning technology should focus primarily on the learners rather than the introduction of new technology (Almohod & Shafi, 2013).

## METHODOLOGY

### Research Design

The researchers applied quantitative research method in conducting the study. This method describes and measures the level of occurrences on the basis of numbers and mathematical calculations (Quantitative Data Collection Methods, 2019). The research provided an analysis of the level of awareness of the students towards the different forms of E–Learning Technology.

Moreover, the product of this research is a technical solution or proposal on how the Campus can improve the current teaching practices by integrating E – Learning platforms. The result of the study can serve as a basis for designing a policy on how to integrate these platforms in the existing learning paradigm in the University.

### Data Collection

The data collection stage of the research focused primarily in gathering information about the implementation of the ICTs in the educational processes of different educational institutions, including the PSU Bayambang Campus. One of the sources of evidence that was employed in the research was interview. Interviews were conducted with the students and faculty members of the campus to determine what are the existing techniques do they employ in delivering instruction.

A survey was also conducted through a questionnaire. The questionnaire is divided into two parts. The first part contains questions about the profile of the subjects including their demographic information, academic background and professional experiences (for faculty members). On the other hand, the second part of the questionnaire contains question that will determine the level of awareness of the subjects towards different e – learning technologies.

Furthermore, secondary data such as literature, ordinances regarding the e–learning implementation and the report of the existing computer facilities as well as the software that are available in the educational institution will be collected using documentation and retrieval of archival records.

## Tools for Data Analysis

After employing data collection methods, data gathered from the subjects were analyzed using the following tools:

**Frequency Counts.** This tool is considered as the most straight forwarded approach in working with quantitative data. Items are classified according to a particular scheme and an arithmetical count is made of the number of items (or tokens) within the text which belongs to each classification (or type) in the scheme (McEnery & Wilson, 1997). In the context of the study, this tool was used to describe the profile of the subjects of the study which included personal information and academic background of the respondents. It was utilized to determine the total percentage of population who are familiar with the different e – learning tools and technologies as well as the features present in these tools. Thus, the following formula will be used.

$$Frequency \% = (number\ of\ occurrence\ /\ total\ population)\ x\ 100\% \qquad Equation\ 1$$

**Average – Weighted Mean.** The researcher will employ AWM to determine the level of awareness of the students towards the e–learning features and technologies. Moreover, the following 5 – point Likert scale was used to interpret the result (Table 1).

Table 1. Five – Point Likert Scale

| SCALE | RANGE | INTERPRETATION |
|---|---|---|
| 1 | 1.00 – 1.49 | Strongly Disagree |
| 2 | 1.50 – 2.49 | Disagree |
| 3 | 2.50 – 3.49 | Undecided |
| 4 | 3.50 – 4.49 | Agree |
| 5 | 4.50 – 5.00 | Strongly Agree |

# RESULTS

## *Profile of the Respondents*

The subjects of the study are the students of PSU – Bayambang Campus during the midyear classes. The students were chosen randomly across different programs and year level. Table 2 shows the distribution of the population of the subjects of the research based on Age, Sex, Program, Year Level and their Academic Status.

Table 2. Distribution of the Population of the Subjects of the Study

| | Frequency | % |
|---|---|---|
| **Age** | | |
| 16 – 23 | 260 | 92.86 |
| 24 – 30 | 19 | 6.79 |
| 31 – 37 | 0 | 0.00 |
| 38 – 44 | 0 | 0.00 |
| 45 – 51 | 1 | 0.36 |
| **TOTAL** | **280** | **100.00** |
| **Sex** | | |
| Male | 127 | 45.36 |
| Female | 153 | 54.64 |
| **TOTAL** | **280** | **100.00** |
| **Programs** | | |
| Bachelor in Elementary Education | 70 | 25.00 |
| Bachelor in Secondary Education | 63 | 22.50 |
| Bachelor of Science in Information Technology | 96 | 34.29 |
| Bachelor of Science in Business Administration | 32 | 11.43 |
| Bachelor in Public Administration | 11 | 3.93 |
| Bachelor of Arts in English Language | 8 | 2.86 |
| Bachelor of Science in Nursing | 0 | 0.00 |
| **TOTAL** | **280** | **100.00** |
| **Year Level** | | |
| First Year | 55 | 19.64 |
| Second Year | 52 | 18.57 |
| Third Year | 99 | 35.36 |
| Fourth Year | 74 | 26.43 |
| **TOTAL** | **280** | **100.00** |
| **Academic Status** | | |
| Regular | 256 | 91.43 |
| Irregular | 24 | 8.57 |
| **TOTAL** | **280** | **100.00** |

The result shows that most of the respondents of the study are between 16 – 23 years old which comprises 92.68% of the total population of the subjects. It was also shown that 54.64% of the respondents are male students. In terms of the program of the respondents, 34.29% of the respondents are currently taking up BS in Information Technology and 35.36% of the population are third year students at the time the study was conducted.

## *Usage of Internet Facility for Learning Purposes*

The study also determined the common activities the subjects usually performed using the internet facility of the campus. The activities identified are in relation to the learning tasks of the students. Table 3 shows the learning activities of the students using internet.

Table 3. Usage of Internet for Learning Purposes

| Usage of Internet for Learning Purposes | Mean | Interpretation |
|---|---|---|
| 1. I use Internet for self – study | 4.27 | Agree |
| 2. I download learning content from Internet | 4.13 | Agree |
| 3. I prefer to read e-book | 3.25 | Undecided |
| 4. I use online library for self-study | 3.51 | Agree |
| 5. I prefer to transfer material through e-mail to my friends, teachers | 3.41 | Undecided |
| 6. I feel satisfied when material is collected from Internet | 3.35 | Undecided |
| 7. I learn many things from Internet through trial-error method | 3.78 | Agree |
| 8. I use different educational blogs for interaction | 3.54 | Agree |
| **AVERAGE WEIGHTED MEAN** | **3.66** | **Agree** |

Based on the responses of the respondents, the result shows that the students use the internet for learning purposes as reflected on the overall Weighted Mean of 3.66 which is interpreted as **"Agree".** However, there are still areas wherein the respondents are not decided on using some of the internet resources as a resource for learning. Most of the respondents are still undecided in using e – books as a learning resource, transferring and collecting learning materials through email. On the other hand, the result also showed that most of the respondents agreed that they use internet for self – study.

# Technological Intervention for Learning Management Practices

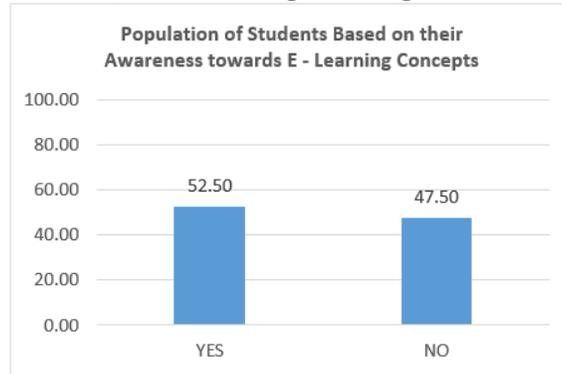

*Figure 1.* Population of Students Based on their Awareness towards E – Learning Concepts

In order to determine the level of awareness of the students of PSU – Bayambang Campus, the researchers also described the population based on their awareness towards the e – learning concepts. Figure 1 shows that 147 students or 52% of the total population are aware of the different e – learning concepts while 48% (133 students) said that they have not yet encountered or come across any e – learning technologies. Hence, the 147 students who are aware of the concept of e – learning will be used as the subject for determining the level of awareness towards the existing e – learning technologies.

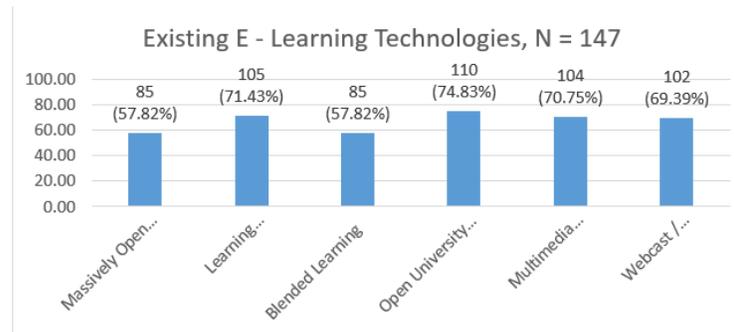

*Figure 2.* Existing E – Learning Technologies

Figure 2 shows the distribution of population (N = 147 students) in terms of their familiarity towards the different e – learning technologies. Some of the most – common e – learning technologies (as used in the study) include Massively Open Online Courses (MOOCs), Learning Management System (LMS), Blended Learning, Open University System, Multimedia Courseware and Webcast Technologies. Based on the result of the survey, it was shown that 74.83% (110 students) of the total number of respondents are familiar with the Open University System. This may be attributed to the fact that the Pangasinan State University is offering programs under Open University System. Aside from Pangasinan State University, universities like University of the Philippines, Polytechnic University of the Philippines and University of the City of Manila (PamantasangLungsod ng Manila) also offers programs under Open University System (Wikipedia,2019).

On the other hand, Massively Open Online Course (MOOC) and Blended Learning are among the e–learning technologies where few students are familiar with. They both comprise 57.82% of the total population which is equivalent to 85 students.

However, it can also be observed that only less than 50% of the total population (280 respondents) are aware of the different forms of e – learning technology. This can be attributed to the fact that the respondents don't have a clear understanding of what constitutes e – learning and thought that it only covers the utilization of computerized instructional materials and slide presentations.

The researchers took the population who are familiar with Learning Management System. Figure 2 shows that there are 105 students who are familiar with different LMS platforms. In line with this, Figure 3 shows some of the LMS platforms along with the distribution of the identified students. Some of these LMS platforms include TalentLMS, Edmodo, Brightspace, Absorb LMS, Saka, Schoology, Blackboard, LitmosLMS, Canvas LMS, Moodle, iSpring Learn, Bridge LMS and Courseplay. They were identified as the top LMS software solution for 2018 (20 Best LMS Software Solutions of 2018, 2018).

The findings showed that among the identified LMS platforms, it was Blackboard which appears to be known among the respondents who were familiar with LMS. This comprises 60.95% of the 105 students or equivalent to 64 respondents. It was followed by Edmodo which has 57 respondents. On the other hand, Sakai was found out to be the least known LMS platform with 7 respondents only.

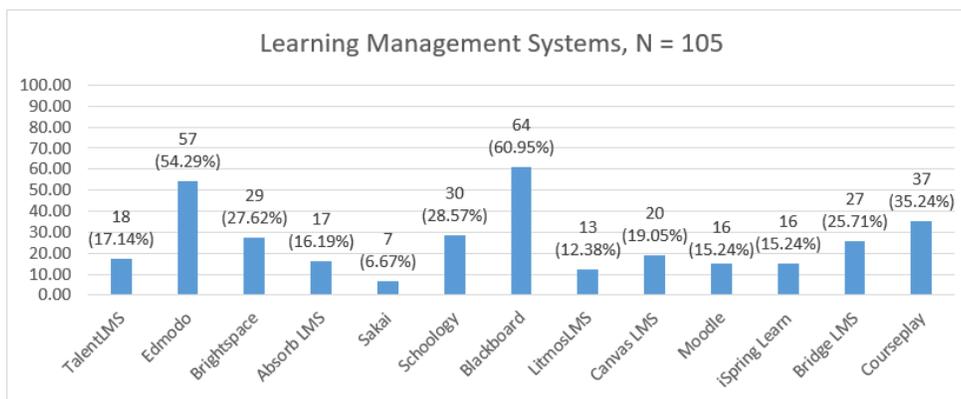

*Figure 3.* Distribution of Population in terms of student's familiarity with different LMS Platforms

Another e – learning technology is the Massively Open Online Courses, popularly known as MOOCs. MOOC is a model for delivering learning content online to any person who wants to take a course, with no limit on attendance (EDUCAUSE, 2018). In the context of this study, the researchers only include six notable MOOC platforms. This includes Cognitive Class, edX, Coursera, Future Learn, Iversity and Udacity. They were included as the top MOOC platform in 2017 (EduTechReviews, 2017).

Among the 147 respondents who appeared to be familiar with e – learning concepts, there are only 85 students who know MOOCs (refer to Figure 2). It appears that Cognitive Class is the most commonly known MOOC platform among others. It was shown that there are 49 students who knows Cognitive Class which comprises 57.65% of the identified population (85 students, refer to Figure 2).It was immediately followed by Future Learn with 47.06% and Coursera with 41.18%. The least commonly known MOOC platform is the Iversity with 23.53% (Figure 4).

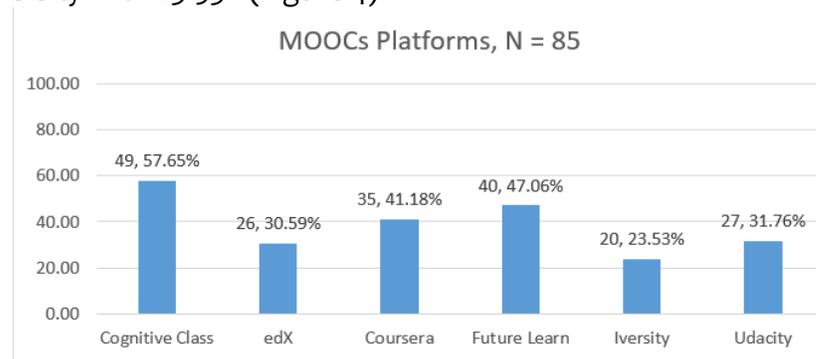

*Figure 4.*Distribution of Population in terms of student's familiarity with different MOOC Platforms

## Known Features of E – Learning Technologies and Frequency of Use

Although e – learning technologies come in different forms, they still share common features. Table 4 shows some of the features present in different e – learning technologies along with the population who were able to use them.

Table 4. Common Features of E – Learning Technologies

|   | Feature | Frequency | % |
|---|---------|-----------|---|
| 1. | Catalogue of Courses / Subjects that we can choose from | 103 | **70.07** |
| 2. | Discussion board for collaborating, sharing experience and exchanging knowledge | 99 | **67.35** |
| 3. | Email Feature for Communication | 91 | **61.90** |
| 4. | Announcement board for putting up messages for all in the class | 91 | **61.90** |
| 5. | An evaluation or assessment tool to check learner progress (Online Quizzes, Exams, etc.) | 92 | **62.59** |
| 6. | A personal notepad for construction of personal knowledge | 81 | **55.10** |
| 7. | A resource center to hold other course related tools and references such as e – books, video lectures, course notes, etc. | 97 | **65.99** |
| 8. | Access to classmates, instructors, experts and technical supports | 96 | **65.31** |
| 9. | View one's progress in a course or subject | 82 | **55.78** |

(Items are adopted from Oye, Salleh & Iahad, 2012; Romiszowski 2004; Latt, Latt & Latt, 2012; Bhatia, 2011); N = 147

The result shows that 103 respondents or 70.07% stated that the e – learning technology they used contains Catalogue of Courses / Subjects that they can choose from. Another feature that seems to be useful when using this technology is the presence of discussion board which is usually intended for collaboration and sharing of experiences and knowledge among its learners. This feature comprises 67.35% of the total population or equivalent to 99 respondents. 65.99% of the respondents also stated that there are intended resource center on the technology they used that gives them access to the learning materials such as e – books, video lectures and course notes. On the other hand, features like accessing one's progress in a course or a subject (82 respondents or 55.78%) and the presence of personal notepad (81 respondents or 55.10%) were the least features known to the respondents.

It appears that there are students who are familiar or even used the different e – learning technologies and its features. However, it is also important to note that being familiar to these features does not correspond to the extent to which the respondents used them. Figure 5 shows the extent of use of the technologies identified and its features among the respondents.

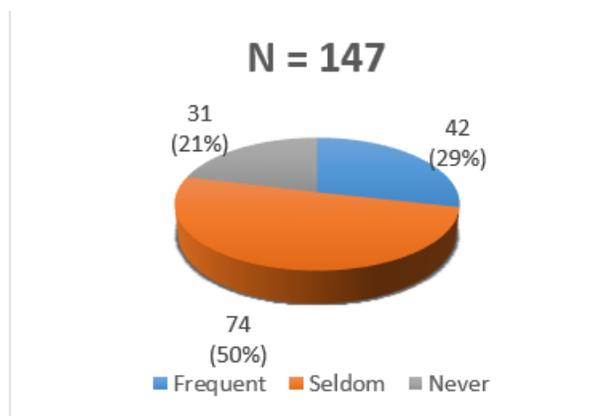

*Figure 5.* Extent of Use of the E – Learning Technologies and its Features

The result shows that 74 respondents or 50% of the population, although they are familiar with the technology, do not always engage themselves in using them. There are only 29% among the respondents who said they use these technologies often and 21% of the population never use them to supplement learning activities. This can be supported by the fact that, in spite of the presence of these technologies to support modern education, the instructors themselves still opt to use traditional methods in teaching and delivering instructional materials to the students. Table 5 shows the different methods that the instructors used in teaching. The respondents were asked to rate the methods from 1 (Never) to 3 (Frequent)

Table 5. Common Methods of Delivering Instruction and Teaching

| Methods | WM | Interpretation |
|---|---|---|
| PowerPoint Presentation | 2.8421 | Frequent |
| Social Media | 2.3609 | Seldom |
| Modules (Lecture Notes, Handouts) | 2.6466 | Frequent |
| Others | 1.1461 | Never |

The table shows the usual methods do the instructors used in delivering instructions to the students. Among these methods, the PowerPoint Presentation and Modules appear to be the most frequently used methods of the instructors while social media are only being used occasionally. However, the result also revealed that aside from these the identified methods; the instructors have not tried using other techniques or technology in teaching.

## ICT Infrastructure of Pangasinan State University – Bayambang Campus

Currently, the Pangasinan State University – Bayambang Campus has 2 Computer Laboratories that caters the computer related courses of the campus. Table 6 provides the summary of the available resources in the computer laboratories.

Table 6. Summary of the Status of Available ICT Resources for ICT Laboratories

| Laboratory Components | Computer Laboratory 1 | Computer Laboratory 2 |
|---|---|---|
| Serviceable Units | 34 | 43 |
| Defective Units | 13 | 7 |
| Internet Connection | None | None |
| Types of Internet Connection Available | None | None |

The table shows the summary of the current status of the laboratory resources available. It can be observed that even though the number of serviceable units is higher than the defective ones, there are no available internet connection in the laboratory which is a primary requirement to access e – learning tools such as LMS and MOOCs. Hence, due to the unavailability of the internet resource inside the laboratory, the students are not given enough exposure to these tools. In addition, the amount of time that they can spend in using the laboratory is very limited since it is also being used as a classroom for many courses. It can be augmented by the availability of internet resources in their homes. Table 7 shows the distribution of population based on the availability of internet connection at home.

The result shows that although there are students who have available internet connection at home, most students have no Internet connection at home (90.36%). These students may have been renting internet services in a computer shop or availing the 30 minutes internet access per day provided by the ISP (Internet Service Provider) of the campus. The amount of time allotted per student in a day is still not enough to support their needs to access internet resources.

Table 7. Distribution of Population based on the availability of the Internet Connection at home.

| Classification | Frequency | % |
|---|---|---|
| Students with Internet Connection at home | 27 | 9.64 |
| Students without Internet Connection at home | 253 | 90.36 |
| **Total** | 280 | 100 |

## DISCUSSION

According to Fitzpatrick (2012), there 5 primary factors / indicators that will determine the success of e – learning technology implementation in a given area. This includes Technology, Design, Human, Support and Evaluation. Technology is considered as the main factor that affects the implementation of the e – learning tools in a given setting. It was followed by the Design which deals with the interface, content and frameworks of a given tool. Next is the Human which is concerned with the attitude, pedagogy and the communication among participants. The fourth one is the Support which has something to do with the levels of training that an organization provides to the users in order for them to use the technology. The last factor is Evaluation. This factor covers the quality and usability of the technology. In the context of this study, the only factors that will be considered include technology, human and support since the current study deals with determining the level of awareness towards the existing e – learning technologies.

**Technology.** Even though there are existing venues where the students can use computers, there is a very limited resource available to enable them to use e – learning platforms. The provision of internet connection is a necessary requirement for them to be able to access these platforms. However, the internet connection provided by the campus is not enough since the students are only given a very limited access time. The amount of time given to the student is not enough to fully use and access the resources available in an e – learning tool. Hence, the inadequacy of the needed infrastructure hinders the student from gaining understanding and knowledge about the different forms e – learning tools.

**Human.** It has been revealed that there are greater number of respondents who are aware of e – learning. However, this does not prove their understanding towards e – learning. During the interview, the respondents thought that e – learning is just a matter of using PowerPoint Presentation, submission of works via email or social media and involvement in a forum. It was also supported by the findings of the study that very few respondents have used different platforms in the area of Learning Management Systems and MOOCs. It only proves that the respondents have a very limited understanding about the concept of e – learning and what are the tools that they can use to implement such approach. Thus, it is better to consider that the organization who wants to implement e – learning in their education system should provide a means on how they can educate the users and expose them to actual practice.

**Support.** It was shown that the respondents have misconception about e – learning concepts. This is due to the fact that the University education system has not yet implemented any forms of LMS in supporting the teaching–learning process or encourages the usage of MOOCs. One of the factors considered important is the support coming from the organization. The organization should find ways on how to educate the users about the concept of e – learning and the benefits that can be derived upon using these tools in education. Aside from providing the needed training, the organization should also enhance the ICT infrastructure that is needed in order to implement the technology and enable its users to maximize its benefits.

## CONCLUSIONS AND RECOMMENDATIONS

The level of awareness cannot be equated with the understanding of concepts about e – learning. The users may know the technology but due to lack of experience in using these technology or inability to access, it may result to the poor understanding and utilization of the e – learning tools available.

There are factors affecting the level of awareness of the students towards e – learning technology. These factors include Support, Technology and Human. Among these factors, it is the Support coming from the organization that has the greatest impact. Trainings can be provided by the organization in order for the users appreciate the concepts and benefits of the e – learning tools. Hence, having better understanding of the technology will increase their willingness to use these tools in improving their technique in delivering and providing instruction. Thus, it will result to the positive attitude of the users towards the technology.

Likewise, exposure to the actual e-learning technology can increase the engagement of the users towards the utilization of the available tools. Hence, the organization should also consider providing the needed ICT infrastructure to support the implementation of E – learning platforms. This includes the provision of additional laboratories solely intended for e – learning activities, providing computers that allow the students to use these tools and the provision of internet connection in the laboratories and the all throughout the campus.

## IMPLICATIONS AND FUTURE WORKS

Some of the highly recognized Universities in the Philippines such as University of the Philippines – Open University System, Ateneo De Manila University, De La Salle University and University of Sto. Tomas, to name a few, have been integrating different Learning

Management Systems to their existing learning process. This integration increases the academic achievements of Filipino College Students among these Universities (Garcia, 2017).

Our study aims to determine the extent of awareness among the students in PSU – Bayambang Campus towards the e – learning platforms. It provides a practical implication to the University that can help in designing programs for educating both the faculty members and students on the benefits of using these technologies which can be applied in the entire PSU Education System. The result of the study can also serve as a baseline for drafting a policy for the integration of Learning Management Systems (or other e – learning platforms) in the existing learning process of the university.

Moreover, since the study is limited to the determination of the level of awareness of the students, future researchers may consider conducting studies which will focus on the comparative analysis of the level of acceptance of the students and faculty members towards e – learning platforms among the Universities and Colleges in Rural and Urban Areas in the Philippines.

## ACKNOWLEDGEMENT

The research was funded and conducted at the Pangasinan State University - Bayambang Campus.